\documentclass[twocolumn,showpacs,preprintnumbers,amsmath,amssymb,prl,tightenlines,final]{revtex4}
\usepackage{hyperref,graphics} 

\usepackage{graphicx}% Include figure files
\usepackage{dcolumn}% Align table columns on decimal point
\usepackage{bm}% bold math
\usepackage{epsfig}
\usepackage{longtable}

\begin{document}

\title{\bf {HfO$_2$: a new direction for intrinsic defect 
driven ferromagnetism}} 
\date{\today} 
\author{Chaitanya~Das~Pemmaraju}
\author{S. Sanvito}
\affiliation{Department of Physics, Trinity College, Dublin 2, Ireland}

\begin{abstract}
In view of the recent experimental reports of unexpected ferromagnetism in  
HfO$_{2}$ thin films \cite{Coey}, we carried out first principles investigations 
looking for magnetic order possibly brought about by the presence 
of small concentrations of intrinsic point defects.  {\it Ab initio} electronic 
structure calculations using density functional theory (DFT) show that isolated 
cation vacancy sites in HfO$_{2}$ lead to the formation of high spin defect 
states. Furthermore these appear to be ferromagnetically coupled with a rather
short range magnetic interaction, resulting in a ferromagnetic ground state for the 
whole system. More interestingly, the occurrence of these high spin states and 
ferromagnetism is in the low symmetry monoclinic phase of HfO$_{2}$. This is radically 
different from other systems previously known to exhibit point defect ferromagnetism, 
warranting a closer look at the phenomenon. 
\end{abstract}


\maketitle

Pure Hafnium oxide (Hafnia, HfO$_{2}$) is a wide band gap insulator with a high dielectric constant 
and presents no evidence of any magnetic order in the ground state. It has been heavily studied in 
recent times within a first-principles context \cite{Medvedeva,Zhao} along with several other 
high-k dielectrics mainly because of its potential for substituting SiO$_2$ as a gate 
dielectric in microelectronic devices. First-principles investigations of the physics of 
defects in Hafnia \cite{Foster,Scopel} also reflect this trend. 

Given the intrinsic non-magnetic nature of HfO$_{2}$, the discovery of ferromagnetic order in 
thin films which are pure except for the possible presence of intrinsic defects \cite{Coey} 
comes as a surprise. Since the nominal valence of Hf in HfO$_{2}$ is 4+, which leaves Hf 
atoms with an empty $d$-shell, the phenomenon was initially termed \textit{$d^0$} magnetism. 
It was suggested that the magnetism probably arose from partially filled $d$-orbitals derived 
from Hafnium atoms co-ordinating Oxygen Vacancy sites ($V_\mathrm{O}$). In this letter we present 
the results of our first-principles band structure calculations which clearly show that 
the observed ferromagnetism is most likely due to the presence of cation (Hf) 
vacancy sites $V_\mathrm{Hf}$. These form high spin states derived mainly from the 
Oxygen $p$-orbitals co-ordinating the $V_\mathrm{Hf}$.

Intrinsic point defect driven ferromagnetism in otherwise non magnetic compounds 
has been previously studied using first principles methods, in several systems. Notable cases 
include CaB$_6$ \cite{Monnier}, CaO \cite{Elfimov} and SiC \cite{Zywietz}. The main 
characteristic of the vacancies in all these systems is the high symmetry, either 
octahedral or tetrahedral around the vacancy site. This invariably leads to a highly 
degenerate single particle spectrum, which may then present high spin states. In CaO 
for instance Coulomb repulsion stabilizes the two holes occupying the degenerate molecular 
orbital associated to $V_\mathrm{Ca}$ in a triplet ground state \cite{Elfimov}.
Similarly Zywietz \textit{et.al} showed that for a Si vacancy in cubic SiC, only the occurrence of a 
magnetic Jahn-Teller distortion stabilizes the spin singlet relative to the triplet 
state otherwise expected from the \textit{T$_{d}$} symmetry and the degenerate single particle 
spectrum. Hafnium vacancies in monoclinic HfO$_{2}$ however, are radically set apart by the 
complete lack of symmetry around the vacancy site. In a defect molecular model, the single 
particle spectrum is completely non-degenerate and yet, as our DFT calculations show, 
they have magnetic ground states. This suggests that our understanding of defect induced 
ferromagnetism is at least incomplete and the phenomenon deserves a closer look. 

Our DFT calculations are carried out in the local spin density approximation (LSDA) using 
the Ceperly-Alder exchange correlation potential \cite{CA} in the numerical implementation
contained in the code {\sc siesta} \cite{siesta}. Test calculations using the generalized 
gradient approximation (GGA), yield qualitatively similar results. 
We use the numerical localized atomic orbital basis set implemented in {\sc siesta} including 
polarized orbitals with an energy shift of 0.01 eV \cite{siesta}. The basis set consists of O 
2$s$, 2$p$ states and Hf 6$s$, 5$d$ and 6$p$ states. A set of two zetas and a polarized orbital are used for every shell
except the 5$d$ for which two zetas suffice.
%%% Stefano
%%% Chaitanya, please indicate which orbitals are considered and how many zetas
%%% End Stefano
$k$-point sampling is done 
by using a fixed $k$-grid cutoff of 15.0\AA\ which is equivalent to a 6x6x6 Mokhorst-Pack mesh 
for the twelve atom monoclinic unit cell. For 2x2x2 supercell containing 96 atoms the same cut-off 
is equivalent to a 3x3x3 Monkhorst-Pack mesh.  The energy cutoff defining the equivalent 
planewave cutoff for the numerical grid is set at a value of 250.0~Ryd. In all cases, 
the systems studied are relaxed until all the forces are smaller than 0.05 eV/\AA. 

\begin{figure}[htbp]
\includegraphics[width=0.45\textwidth]{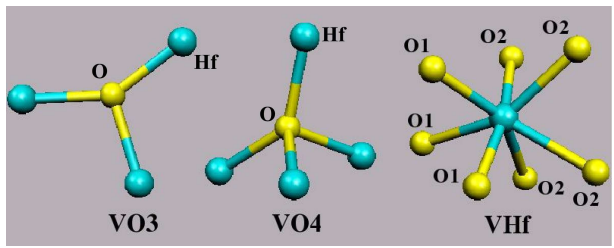}
\caption{\label{vacancies} The geometries around Oxygen vacancies VO3, VO4 and a Hafnium 
Vacancy VHf are shown. The central atoms are included for clarity. The VO4 site forms an 
imperfect tetrahedron while the VO3 site forms an almost planar trigonal structure with 
the co-ordinating Hf atoms.}
\end{figure}

In order to make contact with the \textit{$d^0$} magnetism model \cite{Coey} we first look 
at the case of Oxygen vacancies. In the monoclinic structure there are two
possibilities: i) $V_\mathrm{O}$ co-ordinated by three Hf atoms (VO3), and
ii) $V_\mathrm{O}$ co-ordinated by four Hf atoms (VO4). These are shown in 
Fig.~\ref{vacancies}.  A description of the various defect energetics is outside the scope of 
our work (see reference [\onlinecite{Foster}]) and here we focus only on the electronic 
and magnetic structure of the defects. 

\begin{figure}[htbp]
\includegraphics[width=0.42\textwidth]{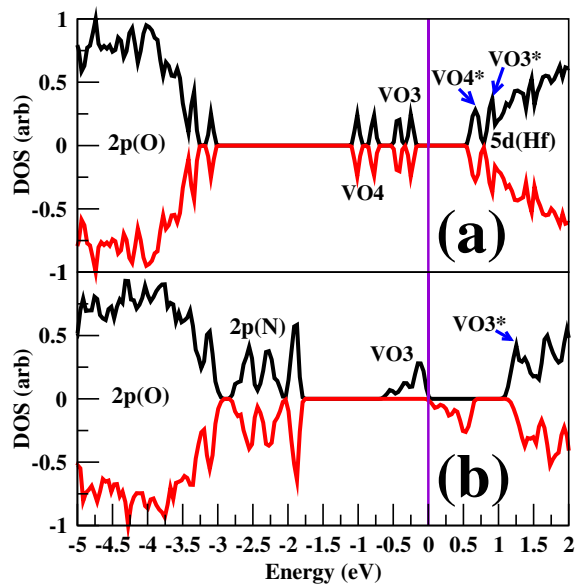}
\caption{\label{2DOVPDOS} Spin resolved total DOS for Oxygen 
deficient HfO$_{2}$. The defect states are labeled according to
figure \ref{vacancies}. VO4* and VO3* denote the empty states at
the bottom of the conductance band. The vertical line indicates
the position of the Fermi level ($E_\mathrm{F}=0$~eV).
(a) Only $V_\mathrm{O}$ are present in the cell, (b) VO3 vacancy co-doped with
N. Note that in this latter case the system is magnetic.}
\end{figure}

An Oxygen vacancy in both the three and four fold co-ordinated case leads to the formation 
of a set of impurity levels, with a low lying level in the HfO$_{2}$ band gap (VO3 and VO4) and 
higher lying levels just below the conduction band (VO3$^*$ and VO4$^*$, see figure~\ref{2DOVPDOS}). 
These impurity levels are formed from the dangling $d$-orbitals of Hf atoms co-ordinating the 
vacancy site. The level in the band gap is filled by two electrons which in the perfect crystal, 
would have populated O $p$-orbitals. In this sense V$_O$ is an n-type defect as 
it results in two electrons occupying conduction band derived cation orbitals.   
Also from the density of states (DOS) it follows that the system is non-magnetic and 
remains effectively semi-conducting. If we assume for the moment that the clusters of Hf 
atoms co-ordinating the VO4 and the VO3 sites form a perfect tetrahedron and a perfect 
trigonal plane respectively, resulting in local \textit{$T_d$} and \textit{$C_{3h}$} symmetries, 
then the single particle molecular orbital ground state in each case would be a 
completely symmetric and non-degenerate \textit{a1} and \textit{$a1^\prime{}$} singlet respectively. 
Similarly the higher lying $\sigma$-bonding single particle states would be a 
\textit{t$_2$} triplet in the VO4 case and a \textit{$e^\prime$} doublet in the VO3 case. 
Deviation from perfect symmetry, as in the actual case, of course means that the degeneracies 
of the excited states are lifted and energy levels suitably reordered. Nevertheless, the 
ground state remains an orbital singlet. 
The higher lying states are well separated in energy from the low-lying singlet 
level. Configurations with one electron promoted to the higher lying states lie 
prohibitively higher in energy relative to the ground state thus ruling out 
electron promotion. Since two electrons occupying an orbital singlet anti-align their 
spins, the resulting ground state is non-magnetic. 

We then investigate the effect of partial p-doping of the system, by substituting an electron 
acceptor like N at an O site. In figure~\ref{2DOVPDOS}b we present the DOS for VO3 only, 
bearing in mind that the situation for VO4 is essentially the same. In this case one electron 
is removed from the $V_\mathrm{O}$ level, which therefore spin splits with a magnetic moment of 
1~$\mu_\mathrm{B}$ per vacancy. We then check whether these localized moments
interact with each other, by doubling our supercell and comparing the total energies of the 
ferromagnetic and anti-ferromagnetic alignment of the magnetic moments on two seperate vacancy sites. 
We find no difference in the total energy and so we conclude that the isolated 
moments are not coupled. The system can at best be paramagnetic and thus the $V_\mathrm{O}$ 
themselves cannot possibly be behind the observed ferromagnetism.  

We now move on to investigate the hypothesis of ferromagnetic order due to $V_\mathrm{Hf}$.
In monoclinic HfO$_2$ (baddeleyite), each Hf atom is co-ordinated by seven O atoms
(see figure~\ref{vacancies}). Of the seven O atoms, three are of one type which we label 
`O1' and the remaining four of a second type which we label `O2'. The two types of O atoms differ in their Hf
co-ordination number in the crystal. We further label the remaining O 
atoms not directly co-ordinating the vacancy site as type `O3'. In the perfect crystal, the O 2$p$ 
levels are fully filled and form the bulk of the valence 
band. Since Hf is a cation with a 4+ valence, the removal a neutral Hf atom introduces four empty 
states among the oxygen 2$p$ levels. The spin occupation of these four
states established whether or not the system is magnetic.  

\begin{figure}[htbp]
\includegraphics[width=0.42\textwidth]{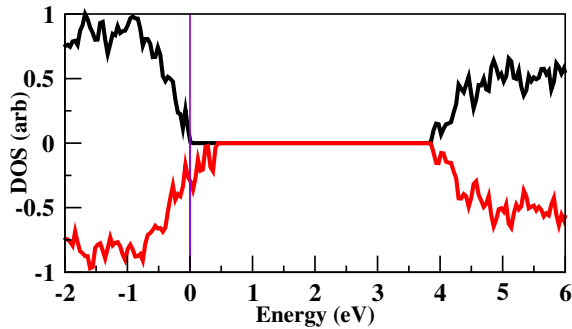}
\caption{\label{HfVac.DOS} Spin resolved DOS for one $V_\mathrm{Hf}$ is a 96 atoms HfO$_2$
supercell. $E_\mathrm{F}$ is set at 0.0~eV.}
\end{figure}

\begin{figure}[htbp]
\includegraphics[width=0.43\textwidth]{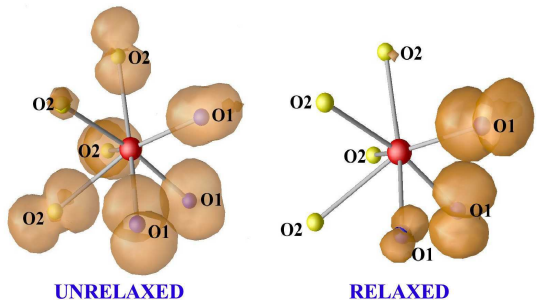}
\caption{\label{MC.VHf.LDOS.2} Isosurface of the local density of states for the highest, 
minority-spin defect level of $V_\mathrm{Hf}$. After relaxation the charge density (hole density) 
clumps mainly on the three O1 type O atoms. As a result the O2 type atoms lose most of their 
spin polarization. The central atom is a dummy atom included only for clarity.}
\end{figure}

\begin{figure}[htbp]
\includegraphics[width=0.42\textwidth]{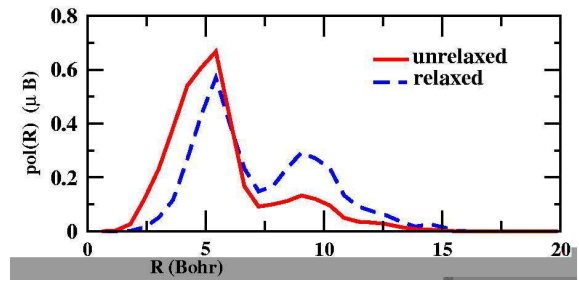}
\caption{\label{mom-loc} Localization plot for the magnetic moment inside the supercell 
before and after relaxation. The quantity presented is $pol(R)=\int_{R}^{R+dR}\mathrm{d}r\int\mathrm{d}\Omega\left[\rho_\uparrow(\vec{r})-\rho_\downarrow(\vec{r})\right]$  against $R$. ${pol(R)}$ is the spin polarization from a shell of 
thickness $dR$ at radius $R$ where $R$ is measured radially outward from the vacancy site.
Note the redistribution of the magnetic moment towards outer shells upon relaxation.}
\end{figure}

At this point we present our LSDA results for the ground state of Hf deficient HfO$_2$. 
We consider a 2x2x2 supercell containing 96 atoms. Only one Hf vacancy is introduced in the
supercell and upon relaxation the O atoms are seen to move outwards around the 
vacancy site by about 0.15~\AA. In figure.~\ref{HfVac.DOS} we present the DOS for the
fully relaxed case. The valence band is clearly spin split, with the compensating holes
mostly confined to the down spin states resulting in an almost half metallic ground state 
with a magnetic moment of 3.52~$\mu_\mathrm{B}$ per vacancy. Prior to relaxation the system 
is completely half metallic with an integer magnetic moment of 4 $\mu_\mathrm{B}$ per vacancy.
The relaxation involves a considerable redistribution of the hole density over 
the O atoms around the vacancy site as shown if figure ~\ref{MC.VHf.LDOS.2}, with the 
magnetism coming predominantly from the O1 type atoms.
The observed charge re-ordering is driven by large scale re-hybridization, upon relaxation, 
of the orbitals constituting the impurity levels with the crystalline surroundings.
We find that the holes clump together on and around the O1 atoms and their nearest O3 type neighbours. 
Figure~\ref{mom-loc} shows the localization of the magnetic moment around the vacancy
site, where this outward spread in the polarization due to relaxation is evident.

\begin{figure}[htbp]
\includegraphics[width=7.5cm]{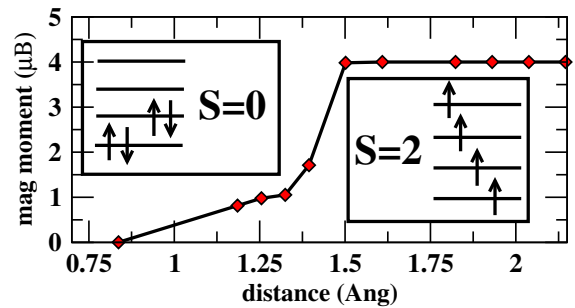}
\caption{\label{Fig6}
Magnetic moment per vacancy as a function of the average distance from the vacancy site of 
the seven co-ordinating Oxygen atoms. In the two insets schematic representations of the
high spin \textbf{S}=2, and non-magnetic \textbf{S}=0 state.}
\end{figure}

Having studied the real space distribution of the magnetic moment, we now look at the symmetry 
aspect of the defect states. Considering the seven O atoms co-ordinating the vacancy site as a 
single molecular cluster, we find a trivial \textit{$C_1$} point group. Thus the single particle 
molecular orbitals generated from combining the O atomic orbitals form a set of non-degenerate 
levels, and the high spin state arises from the singly occupied four topmost molecular orbitals
(see insets of figure \ref{Fig6}).
Interestingly the calculated magnetic moment ({$\sim$4$\mu_\mathrm{B}$}) cannot originate from a single orbitally degenerate
molecular level even in the most symmetric octahedral molecule. In fact the largest orbital 
degeneracy allowed is just three fold, and a total spin \textbf{S}=2 for a configuration of 
four spin 1/2 particles is ruled out by the Pauli principle. This means that in any case
a set of non degenerate orbitals must be involved in the high spin configuration. 
As non-degenerate molecular orbitals differ in single particle properties (kinetic and ionic energies), 
the final configuration of the four spin 1/2 particles is decided not just by mutual Coulomb 
repulsion or exchange energy but by an interplay between all contributions in the Hamiltonian. 

Looking at the relative energetics of different possible electronic configurations of the defect 
levels helps to see what stabilizes the high spin ground state. To this end, we perform fixed 
spin moment calculations, for the two limiting cases of \textbf{S}=2 and \textbf{S}=0.
From the calculations it emerges that the high spin configuration is expensive with regards to 
kinetic and Hartree energies but the gain in exchange and ionic energy is enough to stabilize it.
It is likely that the higher lying molecular orbitals are strongly anti-bonding in character 
with the electrons in these levels being localized more on the ions. This results in higher 
kinetic energies and lower ionic potential energies. In fact, one expects that analogously 
to what happens in molecules, if the O atoms around the vacancy are artificially squeezed 
in towards the vacancy, thus driving the system highly kinetic, the higher lying defect levels 
would be emptied out accompanied by a fall in the magnetic moment. 
This is indeed seen to be the case as shown in figure~\ref{Fig6}. As it 
turns out, at the equilibrium distance in the crystal ({$\sim$2.28$\AA$}) , the magnetic moment is close to 4$\mu_\mathrm{B}$. 
 
Next we address the question of the magnetic coupling between vacancies, by calculating
the total energy of a supercell containing two $V_\mathrm{Hf}$ and comparing the energy
for the ferromagnetic (E$_\mathrm{FM}$) and antiferromagnetic state (E$_\mathrm{AF}$)
(see Table~\ref{Table}).
\begin{table}[htbp]
\centering
\begin{tabular}{c|c}
{\hspace{0.8 cm}$V_\mathrm{Hf}$ concentration\hspace{0.8 cm}} & 
{\hspace{1 cm}$\Delta$E (meV)\hspace{1 cm}} \\
\hline
2.08\% (1) & -113.66 \\
\hline
1.38\% (2) & -50.56 \\
\hline
1.04\% (3) & -13.81 \\
\end{tabular}
\caption{\label{Table}Energy differences $\Delta E=E_\mathrm{FM}-E_\mathrm{AF}$
between ferromagnetic and antiferromagnetic alignment of the magnetic moments 
of two different vacancy sites. The first column indicates the defect 
concentration and the number in bracket the number of Hf sites separating
the two vacancies.}
%%% Stefano
%%%Can you check whether the number between bracket (number of Hf sites separating the vacancies)
%%%is correct?
%%% End Stefano
%%% Chaitanya
%%% Yup, thats correct.
%%% End Chaitanya 
\end{table} 
Clearly the ferromagnetic alignment is always energetically favorable, and most remarkably
the coupling appears rather strong, in particular for short $V_\mathrm{Hf}-V_\mathrm{Hf}$
distances. This leads us to attribute the observed ferromagnetism in HfO$_2$ thin-films \cite{Coey}
to Hf vacancies. Moreover the large values of $\Delta$E suggest Curie temperatures above room 
temperature at large enough concentrations. 

In order to have a better understanding of the origin of this ferromagnetic ordering
we have performed an extensive study of the charge distribution by means of M\"ulliken
population analysis. The main features are: (a) the O atoms in the cell are polarized to 
different degrees depending on their orientation and distance relative to the vacancy site 
but always with the same sign, (b) the Hf atoms in the cell are also polarized 
but importantly, the sign of polarization is opposite to that of the O, and (c) the total 
polarization of all the O atoms in the cell is 3.92$\mu_\mathrm{B}$ and that for the Hf atoms 
is -0.40$\mu_\mathrm{B}$, leaving a moment of the cell of 3.52$\mu_\mathrm{B}$.
This suggests that the magnetic coupling between the O atoms in the cell is mediated by
minority spin electron delocalization across the Hf bridge connecting the O atoms. This applies also 
to O atoms belonging to two different $V_\mathrm{Hf}$ sites. The delocalization is larger when local
moments on the two $V_\mathrm{Hf}$ are ferromagnetically aligned resulting 
in lower Kinetic and Exchange energies relative to the anti-ferromagnetic case.
%Such an
%hole is antiferromagnetically coupled to the high spin state of $V_\mathrm{Hf}$ and the
%overall ferromagnetic state is the stabilized by the partial delocalization of the hole.
%Therefore the mechanism for the ferromagnetism is similar to that proposed for 
%(Ga,Mn)As \cite{SS1}, except that in this case the mediating hole is rather localized at
%the bridging Hf sites.

In summary we have performed DFT calculations investigating the possibility of intrinsic
defect driven ferromagnetism in HfO$_2$. Oxygen vacancies form non-magnetic impurity levels
unless p-type co-dopants are present. However, in this case the magnetic interaction is
negligible, ruling out the hypothesis of d$^0$ ferromagnetism \cite{Coey}. In contrast 
Hf vacancies show a high spin state with an associated magnetic moment of 
$\sim$3.5~$\mu_\mathrm{B}$. These are ferromagnetically coupled via minority spin electron
delocalization across the bridging Hf sites, with a large coupling strength suggesting high
Curie temperatures. Thus a new direction for intrinsic defect ferromagnetism is evident
based on the following facts: 1) symmetry driven orbital degeneracy is not a pre-requisite for the existence 
of a high-spin defect ground state, 2) a set of closely spaced single particle levels together 
with strong exchange might be sufficient for the same, 3) suitable mixing of defect 
states with the crystalline environment can lead to ferromagnetic inter-defect coupling. 
These findings suggest that a wider class of systems, not restricted by symmetry and free 
from the possibility of Jahn-Teller like distortions, might actually be open to intrinsic defect 
ferromagnetism.\\

We thank Prof. J.M.D Coey for useful discussions. 
This work is funded by the Science Foundation of Ireland under the grant SFI02/IN1/I175.

\end{document}